# Coexistence of low and high spin states in $La_{18}Co_{28}Pb_3$

Weiyi Xia[1], Vladimir Antropov[1,2], Yongxin Yao[1,2], Cai-Zhuang Wang[1,2*]
[1]Ames National Laboratory, U.S. Department of Energy, Ames, IA 50011, United States
[2]Department of Physics and Astronomy, Iowa State University, Ames, IA 50011, United States

**Abstract**

The electronic structure and magnetic properties of a newly predicted stable ternary compound $La_{18}Co_{28}Pb_3$ are studied using electronic structure analysis. The ground state of this compound is ferromagnetic, with three positions of nonequivalent magnetic Co atoms. A strong dependence of magnetic properties on volume shows that this system is situated near the point of magnetic instability. A coexistence of high- and low-spin ferromagnetic states as a function of volume near equilibrium was discovered. A corresponding spin tunneling splitting was estimated. The stability of the theoretically predicted magnetic ground state was tested by varying the Hubbard parameter. The thermal spin fluctuations were added to estimate the paramagnetic moment and a Curie temperature. The necessity of experimental verification of the obtained results is emphasized.

Many solid-state systems exist near the point of magnetic instabilities [1]. These instabilities can be related to the appearances of both local magnetic moment and long-range magnetic order [2]. It has been shown that all these systems possess large amounts of spin fluctuations (SF), which can be low-temperature quantum paramagnetic and high-temperature classical (thermal) local moment fluctuations [3,4]. All these fluctuations are essential for determining the right magnetic ground states and the thermal properties of paramagnets and ordered magnets. Significant SF have been observed experimentally in such systems as Invar and antiInvar alloys, magnetoresistive manganates, and numerous paramagnets (like Pd) [5]. A critical consequence of SF for the superconductivity has also been observed in cuprates and iron pnictides [6-7]. The mechanism of SF-induced superconductivity in these materials still needs to be better understood. However, the presence of strong SF appears to be an essential ingredient of such superconductivity. Therefore, searching for systems with strong SF near the point of magnetic instability or quantum critical point (QCP) is a natural strategy for novel superconductor discovery. Such magnetic instability can be both ferromagnetic (FM) and anti-ferromagnetic (AFM) because the coexistence of FM/AFM and superconductivity has been established experimentally. However, the inclusion of SF in the current electronic structure theories needs to be better established. While numerous theories of SF have been proposed in the past, only a few connections between those theories and the physics of realistic materials have been established. The dynamical mean field theory (DMFT) method, while popular, still needs to be revised to the correct description of major figures of merits of metallic magnetism, such as Curie temperature and susceptibility [8].

To search for new materials, one may focus on ternary compounds containing immiscible pairs of elements, and if one of them is magnetic (such as Co-Pb, Ba-Fe, etc.), one can hope to generate SF-rich systems. Still, when such ternary compounds do form, the immiscible elements usually are segregated from each other, with the third element encapsulating or separating them thus leading to reduced dimensionality of the crystal structures [9]. When a *3d*-transition metal (TM) is forced to adopt such reduced dimensionality, complex electronic and magnetic states such as superconductivity [10] and fragile magnetism [11-12] can emerge.

Recently, a ternary $La_{18}Co_{28}Pb_3$ compound containing an immiscible pair of Co and Pb was predicted using machine learning (ML) and first-principles calculation methods [13]. In this study, we investigate the electronic and magnetic properties of this compound. We show that the ground state of this ternary compound is ferromagnetic, with a quantum critical point (QCP) appearing under pressure. Multiple magnetic orderings, including FM and AFM phases, are revealed. The coexistence of high- and low-spin FM states around the equilibrium is predicted. The influence of SF on magnetic properties obtained in the density functional theory (DFT) is studied in the frame of a rotationally invariant form of SF addition using a coupling constant integration formalism [14].

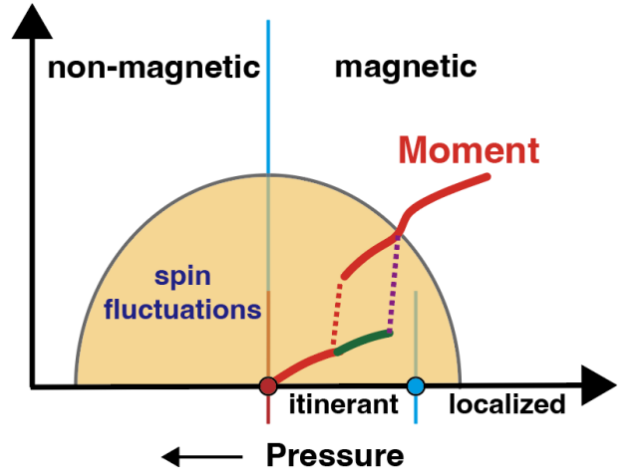

**Fig. 1.** Schematic diagram of spin fluctuation appearance near quantum critical point for the case of non-magnetic-itinerant-localized transition as a function of pressure. The red dot corresponds to the Stoner criterion fulfilling, while the blue dot corresponds to the localized moment criteria fulfillment. An itinerant-localized transition is shown here with the coexistence of two different magnetic solutions.

By SF here, we assume not Hubbard-like Coulomb static correlations but temporal collective variables describing overdamped dynamically related SF, paramagnons, and itinerant fluctuations of spin density not related to the local moment fluctuations, as schematically shown in Fig. 1. These fluctuations are missing in simplified approximations like local density approximation (LDA) or generalized gradient approximation (GGA) for the exchange-correlation energy functionals. In addition, the coexistence of two or more magnetic states leads to the appearance of the spin tunneling effect, which can change the overall stability of the magnetic states. Due to all these limitations, it has been shown that LDA/GGA methods substantially overestimate the tendency to magnetism at T= 0 K for the itinerant magnets (i.e., ZnZr2, NiAl, iron pnictides, etc.) [15-17]. However, both LDA and GGA methods can identify systems near QCP. Below, we will demonstrate that the recently predicted $La_{18}Co_{28}Pb_3$ ternary compound can be considered a system near QCP.

The DFT calculations are performed by using the Perdew-Burke-Ernzerhof (PBE) exchange-correlation functional within the framework of GGA [18] as implemented in the Vienna Ab initio Simulation Package (VASP) [19-20]. A plane-wave cutoff energy is set to 520 eV, and the positions of atoms and the cell parameters are fully optimized. The accuracy of the electron self-consistent field is set to $10^{-4}$ eV, and the Brillouin zone is sampled using a set of gamma-centered uniform 8×8×6 grids. The lattice vectors of the unit cell and the positions of atoms in the unit are fully optimized with the force tolerance of 0.01 eV/ Å.

*Atomic and electronic structures.* The crystal structure of the $La_{18}Co_{28}Pb_3$ compound is shown in Fig 2 (a). It has a tetragonal lattice with an I4/mmm space group symmetry with a complex unit cell containing 98 atoms. There are four Wyckoff sites for La (4c,8i,16m,8j), three Wyckoff sites for Co (32o, 16n, 8h), and two for Pb (2b, 4e), respectively. Within the symmetrized unit cell, the atoms distribute several layers along the **c** direction. For each La layer, the La atoms form an inner ring with the center occupied by the Pb atom and four outer rings, which form multiple tetrahedra with Co o and n sites. The four outer rings are then connected by an octahedron between each other with Co o and h sites. The lattice constant a = 14.02 Å and the c/a ratio of 0.714 are obtained after DFT optimization at the GGA-PBE level. The ground state of this compound is FM and the calculated formation energy is only 1 meV/atom above the known convex hull, suggesting that this compound could be synthesizable by experiment.

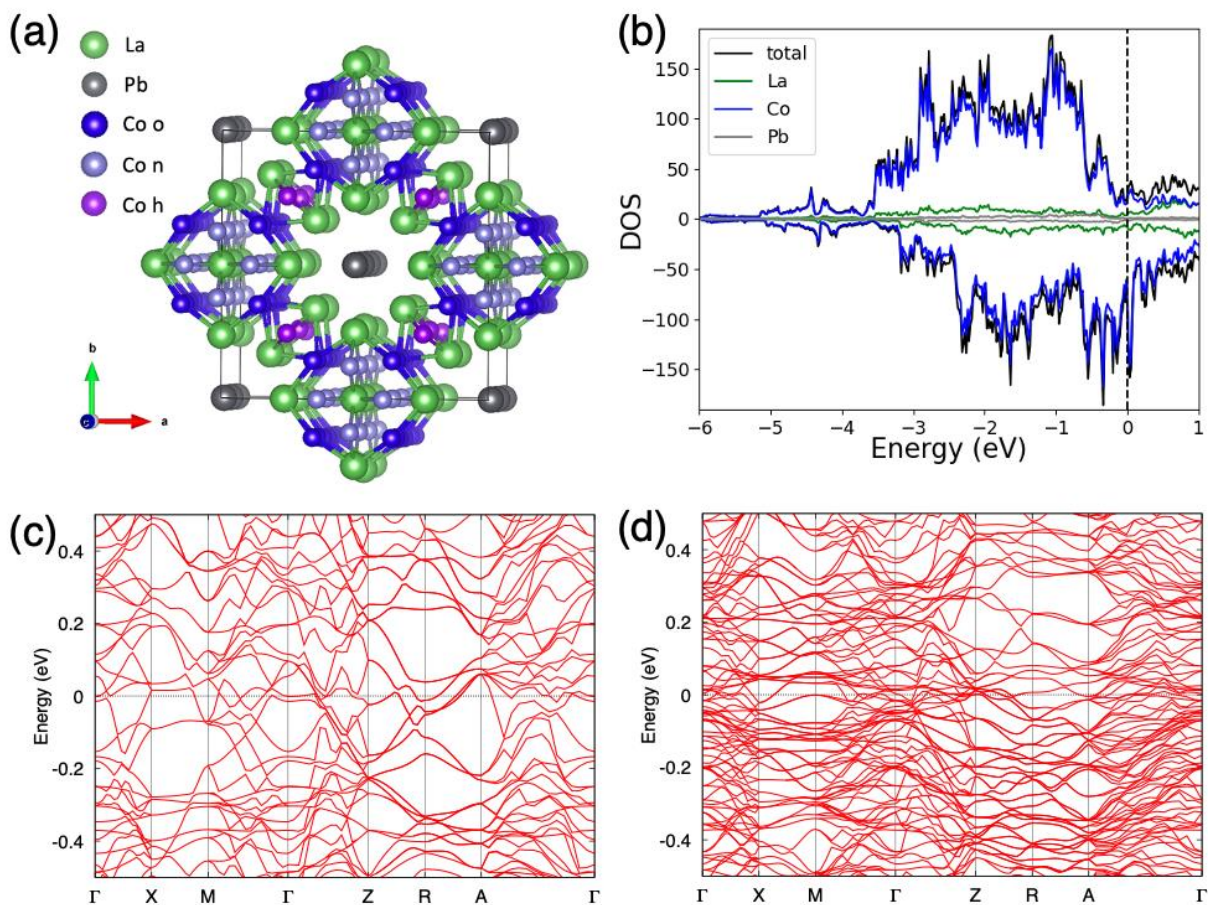

**Fig. 2.** (a) The crystal structure of the predicted $La_{18}Co_{28}Pb_3$ ternary compound. (b) The calculated density of states of the ferromagnetic ground state. We also show the calculated spin-polarized band structure around Fermi level (at 0 eV) for majority spin (c) and minority spin (d).

The total and partial density of states (DOS) is presented in Fig 2 (b). The positive and negative values represent the spin-up and spin-down components of the DOS, respectively. A prominent feature is the high density of states near the Fermi level for the minority spin, primarily originating from Co atoms, particularly the Co o site. The contrasting behavior between the spin-up and spin-down states near the Fermi level is evident in the spin-polarized band structures displayed in Fig 2 (c) for spin-up and Fig 2(d) for spin-down. Notably, the DOS peak near the Fermi level is contributed from some flat bands, as depicted in Fig 2 (c) and (d), representing a Van Hove singularity.

We also calculate the differential electron density for the ground state to investigate the bonding interactions between La, Co, and Pb atoms. In Fig. 3 (a) and (b), the atomic positions and the differential charge density are plotted in the ab-plane cutting through the center Pb atom. The formation of bonding interactions between the Pb atom and surrounding La atoms can be seen. No direct interactions between the Pb and Co atoms are observed. The La atoms connect Pb and Co atoms with La-Pb and La-Co bonds to ensure the formalization and stabilization of this ternary compound involving Pb-Co immiscible pairs.

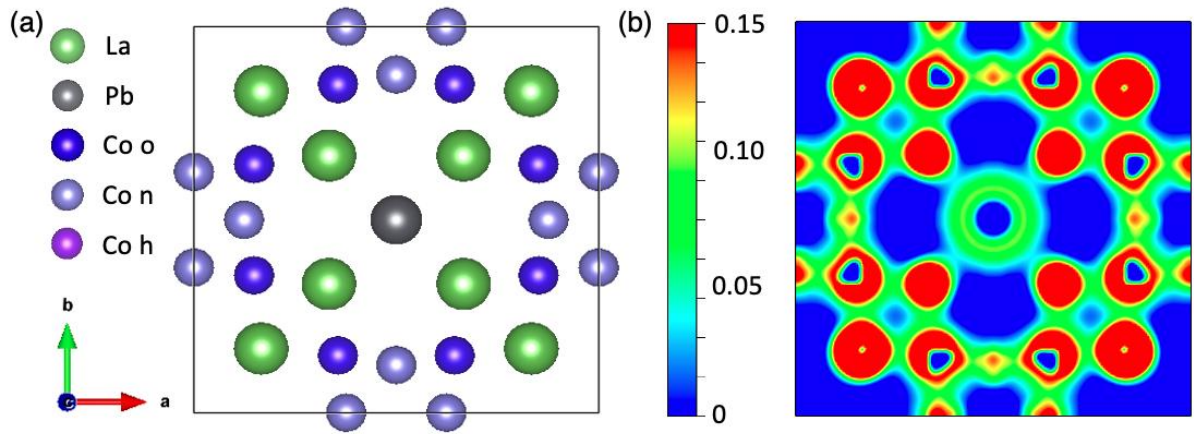

**Fig. 3.** (a) The atomic positions (a) and the differential charge density (b) in the ab-plane cutting through the center Pb atom in the ground state FM $La_{18}Co_{28}Pb_3$ compound.

*Competing magnetic states.* While the ground state of the $La_{18}Co_{28}Pb_3$ compound is FM with an average magnetic moment of 0.63 $\mu_B$ per Co atom (35 $\mu_B$ per unit cell with 2 formula units), our calculations also reveal another FM state with a magnetic moment of only 0.20 $\mu_B$ per Co atom (11 $\mu_B$/cell). We refer to these two FM states as high-spin and low-spin FM states. In addition to the FM configurations, we found many AFM configurations. For simplicity, we investigate three AFM configurations: two a-b in-plane AFM (AFM1 and AFM2) and one out-of-plane AFM (AFM3), as illustrated in Fig. 4 (a). The total energies as the function of volume for these FM and AFM configurations are shown in Fig. 4 (b). The total energies for the two FM states are close, with the energy difference within 10 meV/atom. The energies of the AFM configurations are higher than the high-spin FM state (but < 10 meV/atom). Our results indicate that while the ground state is a high-spin FM state for $La_{18}Co_{28}Pb_3$, there are multiple competitive magnetic states within a small range around equilibrium volume. The energy difference between these states is within 10 meV/atom range.

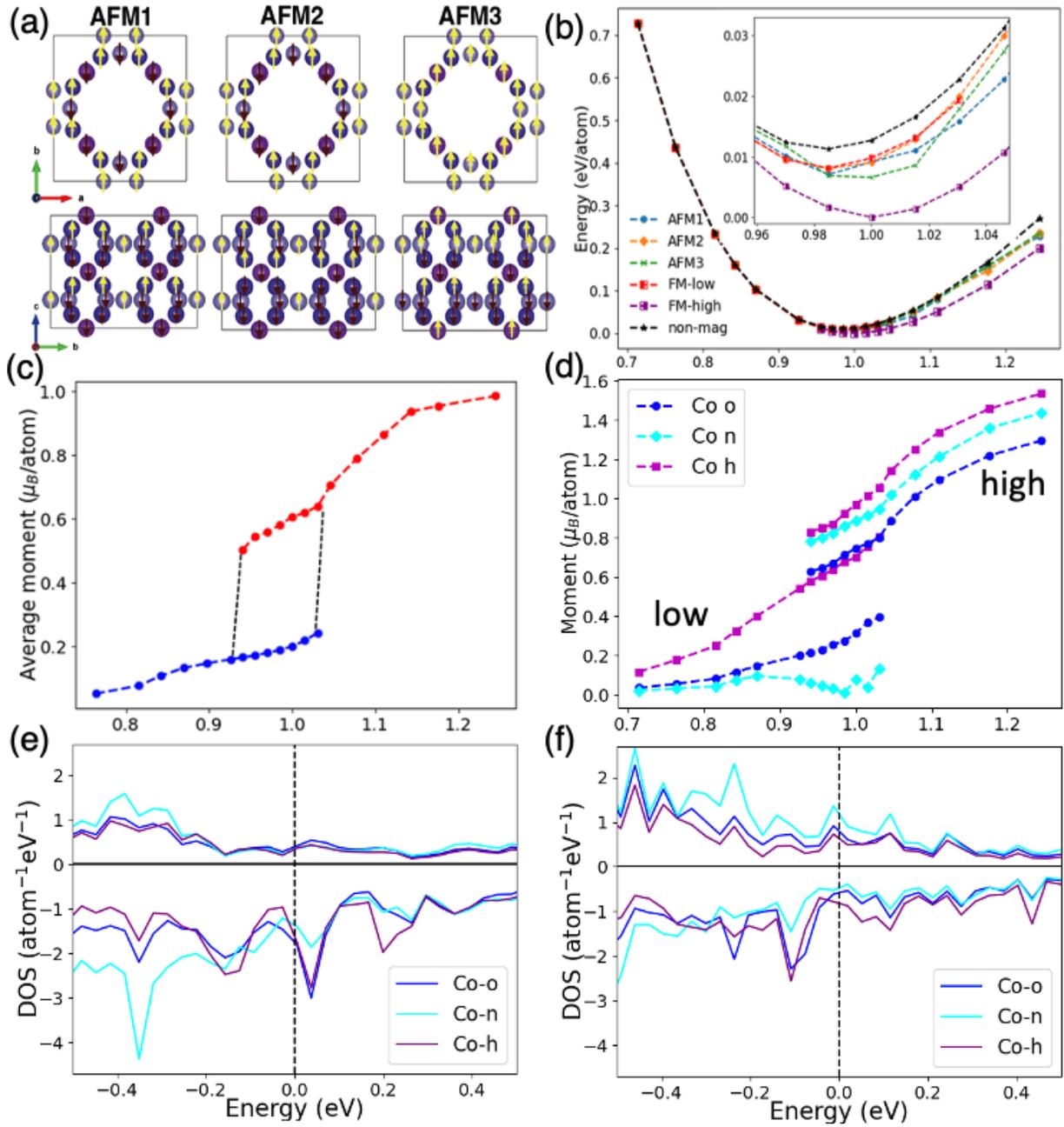

**Fig. 4.** (a) Illustration of three AFM configurations. Left: a-b in-plane AFM configuration (AFM1); middle: another a-b in-plane AFM configuration (AFM2); Right: out-of-plane AFM configuration (AFM3). (b) The total energy as the function of volume change ($V/V_0$) for various states. (c) The averaged magnetic moment per Co atom in the two FM phases in $La_{18}Co_{28}Pb_3$ calculated at different $V/V_0$. (d) The atomic decomposed magnetic moments for the non-equivalent Co atoms at different $V/V_0$. (e-f) The decomposed DOS per atom for the non-equivalent Co atoms in high-spin (e) and low-spin state (f) at the equilibrium volume $V_0$.

*Magnetic phase transition.* Now, we study the volume dependence of the magnetic moment for the low- and high-spin FM states, as shown in Fig. 4 (c). The average magnetic moment per Co atom changes significantly from 0.2 to 0.6 $\mu_B$ around the equilibrium volume. The magnetic moments for the three types of non-equivalent Co atoms at different volumes are also shown in Figure 4 (d). For Co h sites, the magnetic moment does not change much going from low to high spin state, while for n sites, a significant change is observed.

The hysteresis-like loop for magnetization around the equilibrium volume in Fig. 4 (c) and (d) is obtained by reducing (increasing) the volume from the high (low)-spin state in a small increment and using the wavefunction from the previous volume as the starting wavefunction for the self-consistent DFT calculations. Our calculations suggest that the low- and high-spin states in $La_{18}Co_{28}Pb_3$ should be stable separately at a volume below 0.96 $V_0$ and above 1.04 $V_0$, respectively. In the volume range of 0.96 to 1.04 $V_0$, both these states coexist. Fig. 4 (b) also shows the presence of other metastable collinear states in this region of volumes. The existence of such a hysteresis-like loop for magnetization indicates a possible low- to high spin phase transition under pressure.

By comparing the DOS near the Fermi level for high- and low-spin states, one can see a significant difference between two types of the non-equivalent Co atoms, o and n sites, as shown in Fig. 4 (e) and (f). Moving from the low- to the high-spin state, we observe a significant increase in the DOS of Co o and n atoms at the Fermi level for the minority and a significant decrease for the majority spin. The main peak position of the minority spin for all the non-equivalent Co sites shifts from unoccupied states to occupied states, indicating a significant change in the on-site exchange interactions for these Co d orbitals. These observations agree with Fig. 4 (d) that the major difference comes from Co o and n sites.

Next, we study the shape of the adiabatic energy surface between the obtained spin states of $La_{18}Co_{28}Pb_3$ by employing constraining moment calculations. In Fig. 5, the possible path between the low- and high-spin states is shown as a function of the total moment. The constrained magnetic moment calculations are performed by adopting the scheme from [21] implemented in VASP. In particular, the initial moments on each non-equivalent Co site are confined along the out-of-plane direction, the easy direction with the lowest energy. At each constraint as shown in Fig. 5, the ratios of the moments on the three non-equivalent Co sites are kept the same as those at the equilibrium volume for the low-spin (blue) or high-spin (red) calculation, respectively. The weight $\lambda$ of the penalty function is set to 20 for all calculations, and the calculated total energies are converged within $10^{-5}$ eV per magnetic atom. The Wigner-Seitz radius of each atom is chosen so that the sum of the volume of the spheres is equal to the volume of the cell. With such a simple static constraint, we found two FM states with the transition around the averaged moment of 0.3 $\mu_B$. The magnetic adiabatic surface resembles an asymmetric double potential with very low barrier energy (< 1 meV/atom) from the low-spin to the high-spin state. The existence of such potential in the "classical" spin energy would lead to the appearance of the quantum spin tunneling effect.

We estimated this tunneling by employing the ab initio spin instanton technique [22] implemented in the TB-LMTO method.

The tunneling splitting is written as $\Delta(\tau) = \left(\frac{2\hbar\omega_0 T_0}{\pi}\right) e^{\omega_0 \tau - 2S_M}$, where $\hbar\omega_0$ is the initial time dependence of the kinetic energy, $S_M$ is a magnetic contribution to the effective action, $\tau$ is a time to reach the low-spin state, $T_0$ is the initial kinetic energy (used as a trial value). Starting from the high-spin initial configuration, spin instanton dynamics was performed for ten different $T_0$ in the range 0.001- 0.1 meV. For each value of $T_0$, the corresponding initial spin deviation angle was determined by trial and error. Time steps of 0.11 a.u. were used. The resulting splittings appear to be in the range of 0.08 – 0.1 meV/atom and cannot affect the relative stability of the high-spin state.

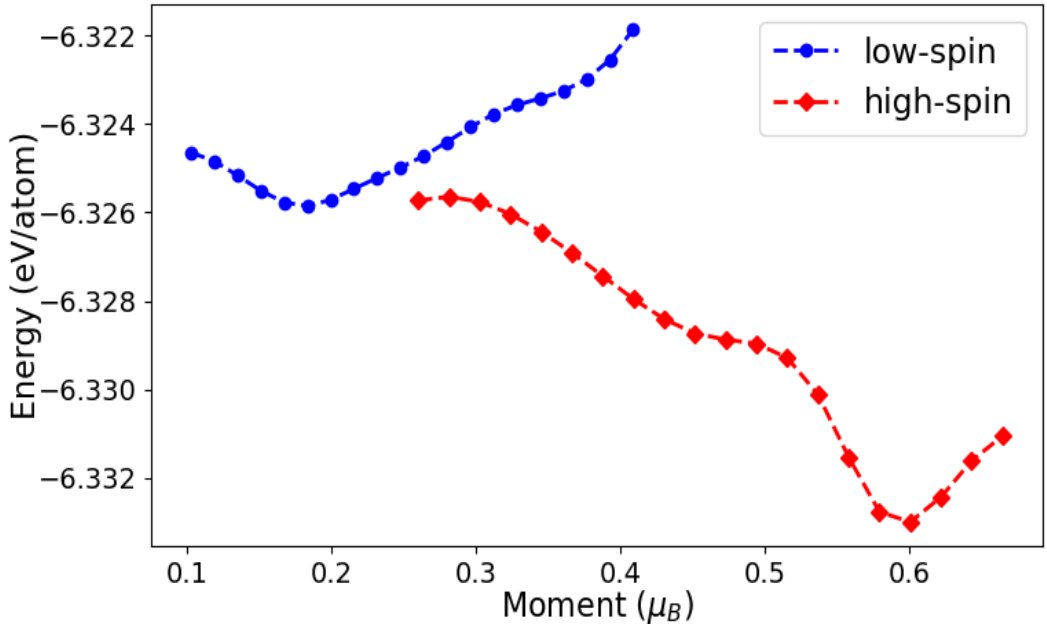

**Fig. 5.** Asymmetric double well potential energy surface between low- and high-spin states in $La_{18}Co_{28}Pb_3$. The directions of the initial moments on each non-equivalent Co site are constrained in out-of-plane direction in the calculation, and the ratios of these moments are kept the same as those at the equilibrium volume, for the low-spin (blue) or high-spin (red) calculation, respectively.

It is worth noting that the DOS of the ground state plotted in Fig. 2 (b) shows that the Co d-states in this compound exhibit a somewhat narrow bandwidth of ~ 3 eV compared to 4 - 4.5 eV in the elemental hcp Co. This could suggest somewhat stronger electronic correlations in this system. We check the sensitivity of the calculated electronic structure and magnetic properties of $La_{18}Co_{28}Pb_3$ to the variations of possible atomic Coulomb correlations by performing DFT+U calculations. Adding the U term on Co atoms in DFT-PBE calculations, we found the magnetic moments of the Co atoms increase significantly and almost linearly with the value of U in the range of 0 - 4 eV (Fig. 6). Whether DFT+U should be used, and which U value should be used in

DFT+U calculation for accurate description of magnetization in this compound remains an open question for future investigations.

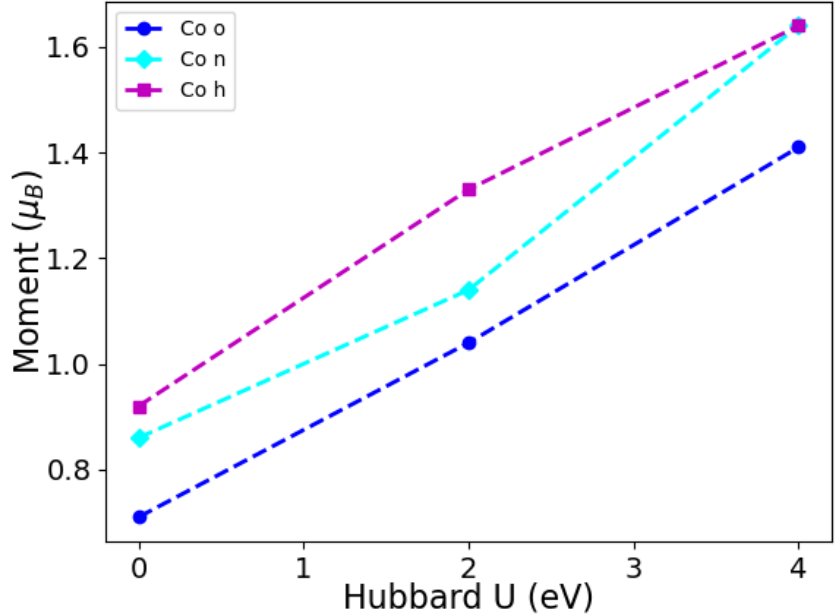

**Fig. 6.** The calculated magnetic moments for different Co sites in La$_{18}$Co$_{28}$Pb$_3$ as a function of the Hubbard U parameter on Co atoms.

*Stoner and Curie temperatures.* Next, we evaluate the Stoner temperatures of the low- and high-spin FM magnetic phases of the La$_{18}$Co$_{28}$Pb$_3$ compound. Fig. 7 shows the unit cell's magnetization as the electronic temperature function for the two FM phases. The Stoner temperature is defined as the electronic temperature above which the magnetic moment of the compound vanishes. From Fig. 7, we can see that the Stoner temperature of the low-spin FM phase is about 900 K. In contrast, the high-spin FM phase can be as high as 4000 K. These results suggest that the Stoner model is unsuitable for our system's localized high-spin FM state. Instead, the Heisenberg model could be more relevant in this case. Within the framework of the Heisenberg model, the Curie temperature $T_c$ can be estimated using the mean-field approximation, which involves calculating the energy difference between the FM and AFM configurations. The mean field expression $T_c \sim \frac{2}{3} J_0$, where $J_0$ is the effective exchange interaction parameter evaluated from the difference $E_{AFM} - E_{FM}$. The $T_c$ for the high-spin FM state appears to be around 60 K, supporting the possible coexistence of two FM states below 60 K.

Alternatively, we can estimate the Curie temperature from the paramagnetic (high temperature) limit. To do this, we first obtain the free energy of magnetic systems with thermal spin fluctuations due to Bose-Einstein excitations (absent in T = 0 K, DFT) using the coupling constant integration [14,17,23-25] of a spin-disordered state. The details of the susceptibility calculations can be found in [26]. It is clear from Fig. 7 that we can safely calculate spin paramagnetic susceptibility below 2000 K for the high-spin state where magnetic moments on the Co atom still look localized enough. For such high-temperature cases, we obtain the susceptibility

$$\chi(\boldsymbol{q}) = \frac{\chi_0(\boldsymbol{q})}{1-(I-\lambda)\chi_0(\boldsymbol{q})} \quad . \quad (1)$$

where for Moriya-Kawabata parameter

$$\lambda(T) = 3kT \int_0^I dI \sum_{\boldsymbol{q}}[\chi(\boldsymbol{q}) - \chi_0(\boldsymbol{q})] \quad (2)$$

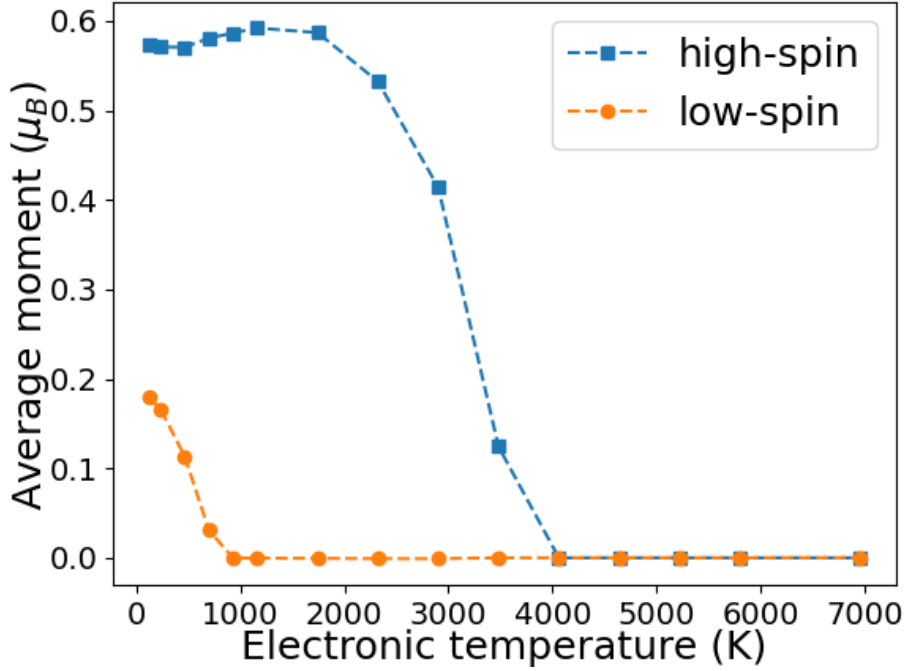

**Fig. 7.** The calculated Stoner temperature of both high-spin and low-spin FM states of $La_{18}Co_{28}Pb_3$.

Our calculations indeed revealed linear dependence of inverse susceptibility below 1300K, which can be fitted with the traditional expression,

$$\chi(\boldsymbol{q} = 0) = \frac{\mu_{eff}^2}{3(T-T_c)} \quad (3)$$

to extract the high-temperature effective moments. For the average effective moment $\mu_{eff}$, we obtained a value of 1.8 $\mu_B$ with the critical temperature of magnetic phase transition $T_c$ = 145 K. We believe this estimate is more reliable than the mean-field values obtained above due to the presence of large itineracy.

We also note that the degree of localization of the magnetic moment and the Curie temperature is strongly dependent on the amplitude of the moment. As shown in Fig. 6, the magnetic moments of the Co atoms increased when the Hubbard parameter U was added. Therefore, the estimated Curie temperature is also subject to change with U. Our DFT+U calculation using U = 2 eV produces a much higher mean field estimate $T_c$ of about 540 K with $\mu_{eff} = 2.1\ \mu_B$. We expect an even larger increase of $T_c$ for larger U values, but such strong correlations could be unlikely for this system. In any case, all obtained numbers, and especially the strong sensitivity of the results to the value of the Hubbard parameter in this metallic system, should be easily verified experimentally when samples are available.

*Summary.* In summary, we investigated the electronic and magnetic properties of the recently predicted $La_{18}Co_{28}Pb_3$ compound containing an immiscible pair of elements Co and Pb. The DFT-PBE calculations predicted an FM state as the ground state. Our study also revealed the coexistence of high- and low-spin magnetic states in the extensive range of volumes that can be accessed experimentally. The spin tunneling effect between these two states using the spin instanton technique appears to be relatively small and cannot influence the stability of the ground state. The calculations of the temperature-dependent susceptibility using integration over the parameter method produced a Curie temperature of about 145 K. We further predicted a strong dependence of the Curie temperature and the paramagnetic moment on the strength of Hubbard type of correlations. Further experimental research is needed to verify our predictions.

**Acknowledgments**

We are grateful to Prof. Paul Canfield for his insightful suggestion on exploring the structures and properties of ternary La-Co-Pb compounds which contain a pair of immiscible elements Co and Pb, and to Dr. Tyler Slade for discussions and his great effort in experimental synthesis of La-Co-Pb ternary compounds. We also acknowledge insightful discussions with Dr. Feng Zhang on theoretical calculations. Work at Ames Laboratory was supported by the U.S. Department of Energy (DOE), Office of Science, Basic Energy Sciences, Materials Science and Engineering Division, including a grant of computer time at the National Energy Research Supercomputing Center (NERSC) in Berkeley. Ames Laboratory is operated for the U.S. DOE by Iowa State University under contract # DE-AC02-07CH11358.

**Data Availability**
The data that support the findings of this paper, including both experimental and computational results, are available from the corresponding author, Cai-Zhuang Wang (email: wangcz@ameslab.gov), upon reasonable request.

**Code availability**
All codes used in this work are accessible through their websites. We use VASP 5.4.1 version in this work.

**Ethics declarations**
**Competing interests**
The authors declare no competing interests.